\documentclass[12pt]{iopart}

\usepackage{bm}
\usepackage{amsfonts}
\usepackage{iopams}
\usepackage{braket}
\usepackage{longtable}
\usepackage{graphicx}
\usepackage{ntheorem}
\usepackage{stackrel}
\usepackage{braket}
\usepackage{cite}
\usepackage{lipsum}

\newtheorem{remark}{Remark}
\newtheorem{lemma}{Lemma}
\newtheorem{prop}{Proposition}
\newtheorem{coro}{Corollary}

\newcommand{\dd}{\,\mathrm{d}}

\begin{document}
\title{Average capacity of quantum entanglement}
\author{Lu Wei}
\address{Department of Computer Science, Texas Tech University, TX 79409, USA}
\ead{luwei@ttu.edu}
\vspace{9pt}
\begin{indented}
\item[]May 2022
\end{indented}

\begin{abstract}
As an alternative to entanglement entropies, the capacity of entanglement becomes a promising candidate to probe and estimate the degree of entanglement of quantum bipartite systems. In this work, we study the typical behavior of entanglement capacity over major models of random states. In particular, the exact and asymptotic formulas of average capacity have been derived under the Hilbert-Schmidt and Bures-Hall ensembles. The obtained formulas generalize some partial results of average capacity computed recently in the literature. As a key ingredient in deriving the results, we make use of recent advances in random matrix theory pertaining to the underlying orthogonal polynomials and special functions. Numerical study has been performed to illustrate the usefulness of average capacity as an entanglement indicator.
\end{abstract}

\vspace{2pc}
\noindent{\it Keywords}: quantum entanglement, entanglement capacity, Hilbert-Schmidt ensemble, Bures-Hall ensemble, random matrix theory, special functions


\maketitle

\section{Introduction}
Crucial to a successful exploitation of advances of the quantum revolution is the understanding of quantum entanglement. Entanglement is the physical phenomenon, the medium, and, most importantly, the resources that enable quantum technologies. Estimating the degree of entanglement over different models of generic (random) states has been a subject of intense study in the past decades. Existing results mainly focus on entanglement entropy based estimation using, for example, von Neumann entropy~\cite{Page93,Foong94,Ruiz95,VPO16,Wei17,Wei20,HWC21,SK19,Wei20b,Wei20c,BHK21,BHKRV21,HW22}, quantum purity~\cite{Lubkin78,Giraud07,Sommers04,Osipov10,Borot12,SK19,Wei20b,LW21}, and Tsallis entropy~\cite{MML02,Wei19T} as entanglement indicators. These results concern, in the setting of bipartite systems, the statistical behavior of entanglement entropies over generic state models of the well-known Hilbert-Schmidt ensemble~\cite{Lubkin78,Page93,Foong94,Ruiz95,MML02,Giraud07,VPO16,Wei17,Wei19T,Wei20,HWC21} and Bures-Hall ensemble~\cite{Sommers04,Osipov10,Borot12,SK19,Wei20b,Wei20c,LW21}, as well as the emerging fermionic Gaussian ensemble~\cite{BHK21,BHKRV21,HW22}. Besides entropies, another type of entanglement indicator is the capacity of entanglement, which was proposed in~\cite{Yao10} as the analog of heat capacity of a thermal system. Entanglement capacity complements entropies in characterizing the entanglement, where distinct behavior between the two types of entanglement indicators has been numerically observed in~\cite{Nandy21}. In quantum information theory, capacity is directly related to fidelity susceptibility and quantum Fisher information~\cite{Boer19}. It is also a useful quantity to diagnose phase transitions relevant to quantum field theory~\cite{Okuyama21}. In particular, the limiting capacity is identified as a critical value to distinguish integrable systems from chaotic systems~\cite{BNP21}. Despite the importance of entanglement capacity, results in the literature are rather limited. Under the Hilbert-Schmidt ensemble, the average capacity for small dimensional systems have been obtained in~\cite{Boer19}, whereas the average of a related but mathematically simpler notion of annealed capacity~\cite{Okuyama} has been computed in~\cite{Okuyama21}. Special cases of the average capacity under the fermionic Gaussian ensemble have been studied recently in~\cite{BNP21,HW22}.

To complete the picture in understanding the typical behavior of entanglement capacity, we compute the average capacity over the major models of generic states. Our main results include the exact yet explicit formulas of average capacity over the Hilbert-Schmidt and Bures-Hall ensembles. We also derive the corresponding limiting values of capacity when the dimensions of the two subsystems approach infinity. The results of this work are obtained by making use of recent progress on the underlying random matrix ensembles as well as the associated orthogonal polynomials and special functions. Numerical simulations are performed to verify the derived results and to show the usefulness of average capacity as an entanglement indicator.

The rest of the paper is organized as follows. In section~\ref{sec:result} we formulate the problem of interest before presenting the main results. Particularly, the exact capacity formulas under the Hilbert-Schmidt ensemble and Bures-Hall ensemble are summarized respectively in proposition~\ref{p1} and proposition~\ref{p2}, where the asymptotics of capacity are given in corollary~\ref{c}. Section~\ref{sec:proof} is devoted to the proofs of the main results. We summarize the key findings and outline potential future works in section~\ref{sec:con}. Relevant summation identities are listed and discussed in appendix A.

\section{Problem formulation and main results}\label{sec:result}
Before discussing entanglement indicators of entropies and capacity, we outline the density matrix formulism~\cite{BZ17}, introduced by von Neumann, that gives rise to the concept of bipartite systems and generic states. Consider a composite quantum system consisting of two subsystems $A$ and $B$ of Hilbert space dimensions $m$ and $n$ (with $m\leq n$), respectively. A generic state of the bipartite system is written as a linear combination of the random coefficients $c_{i,j}$ and the complete bases of subsystems $A$ and $B$ as
\begin{equation}\label{eq:state}
\Ket{\psi}=\sum_{i=1}^{m}\sum_{j=1}^{n}c_{i,j}\Ket{i_{A}}\otimes\Ket{j_{B}},
\end{equation}
where the coefficients $c_{i,j}$ follow independent and identically distributed standard complex Gaussian random variables. The corresponding density matrix of the full system is
\begin{equation}\label{eq:rho}
\rho=\Ket{\psi}\Bra{\psi}
\end{equation}
with the natural probability constraint
\begin{equation}\label{eq:del}
\tr(\rho)=1.
\end{equation}
As opposed to a deterministic state of fixed coefficients $c_{i,j}$, the generic states are ensembles of random states useful in probing and estimating the holistic statistical performance of a system.

The defining feature of the bipartite system is the operation of partial trace (of the full density matrix) leading to a reduced density matrix that models the entanglement between the two subsystems. Different models of generic states are specified by different ways the partial trace is taken. If we directly take the partial trace over the density matrix~(\ref{eq:rho}) of the larger system $B$,
\begin{equation}\label{eq:rdHS}
\rho_{A}=\tr_{B}(\rho),
\end{equation}
the resulting eigenvalue distribution\footnote{We only concern the eigenvalue distribution of a reduced density matrix as the observables of interest, such as entropies and capacity, are functions of the eigenvalues instead of eigenvectors.} (entanglement spectrum) of the reduced density matrix~(\ref{eq:rdHS}) is known as the Hilbert-Schmidt ensemble~\cite{Page93,BZ17}
\begin{equation}\label{eq:HS}
f_{\rm{HS}}\left(\bm{\lambda}\right)\propto
\delta\left(1-\sum_{i=1}^{m}\lambda_{i}\right)\prod_{1\leq i<j\leq m}\left(\lambda_{i}-\lambda_{j}\right)^{2}\prod_{i=1}^{m}\lambda_{i}^{\alpha},
\end{equation}
where $\delta(\cdot)$ is the Dirac delta function and
\begin{equation}\label{eq:a}
\alpha=n-m
\end{equation}
denotes the dimension difference of the two subsystems. The Bures-Hall ensemble is a variant of the Hilbert-Schmidt ensemble in that its state is a superposition of the state~(\ref{eq:state}) as
\begin{equation}\label{eq:BHs}
\Ket{\varphi}=\Ket{\psi}+\left(\mathbf{U}\otimes\mathbf{I}\right)\Ket{\psi},
\end{equation}
where $\mathbf{U}$ is an $m\times m$ unitary random matrix with the measure proportional to $\det\left(\mathbf{I}+\mathbf{U}\right)^{2\beta+1}$~\cite{SK19}. We now take the partial trace over the new density matrix $\rho=\Ket{\varphi}\Bra{\varphi}$, the entanglement spectrum of the reduced density matrix $\rho_{A}=\tr_{B}(\rho)$ is the Bures-Hall ensemble~\cite{Sommers04,BZ17}
\begin{equation}\label{eq:BH}
f_{\rm{BH}}\left(\bm{\lambda}\right) \propto \delta\left(1-\sum_{i=1}^{m}\lambda_{i}\right)
\prod_{1\leq i<j\leq m}\frac{\left(\lambda_{i}-\lambda_{j}\right)^{2}}{\lambda_{i}+\lambda_{j}}\prod_{i=1}^{m}\lambda_{i}^{\beta},
\end{equation}
where
\begin{equation}\label{eq:b}
\beta=n-m-\frac{1}{2}.
\end{equation}
The two ensembles~(\ref{eq:HS}) and~(\ref{eq:BH}) are supported in the set
\begin{equation}\label{eq:D}
\left\{0\leq\lambda_{m}<\ldots<\lambda_{1}\leq1,~~\sum_{i=1}^{m}\lambda_{i}=1\right\},
\end{equation}
which, in particular, reflects the probability conservation~(\ref{eq:del}). In principle, one may construct various other models of generic states in the space of density matrices~\cite{BZ17}. The main reasons that the considered ensembles~(\ref{eq:HS}) and~(\ref{eq:BH}) stand out as the most important ones are discussed below.
\begin{itemize}
\item \textbf{The Hilbert-Schmidt ensemble} corresponds to the simplest model of generic quantum states, where no prior information of the states needs to be assumed. The randomness of the states comes from the assumption of Gaussian distributed coefficients, which correspond to the most non-informative distribution. Namely, the Hilbert-Schmidt ensemble can be thought of as the baseline ``Gaussian model'' universal in statistical modelling of an unknown variable. Therefore, in the investigation of any quantum information processing task, it is always desirable to make use of generic Gaussian states to benchmark the performance.

\item \textbf{The Bures-Hall ensemble} is an improved variant of the Hilbert-Schmidt ensemble that satisfies a few additional properties~\cite{BZ17}. It is the only monotone metric that is simultaneously Fisher adjusted and Fubini-Study adjusted. The Bures-Hall metric, related to quantum distinguishability, is known to be the minimal monotone metric. It is also a function of fidelity, which is a key performance indicator in quantum information processing. In addition, the Bures-Hall ensemble enjoys the property that, without any prior knowledge on a density matrix, the optimal way to estimate the density matrix is to generate a state at random with respect to this measure~\cite{Sommers04}. As a result, it is often used as a prior distribution (Bures prior) in reconstructing quantum states from measurements. It is also known that the generic states from the Hilbert-Schmidt and Bures-Hall ensembles are physical in that they can be generated in polynomial time~\cite{Oliveira07}.
\end{itemize}

Entanglement serves as a measure of the non-classical correlation between the subsystems $A$ and $B$ of a bipartite system. The degree of entanglement can be estimated by entanglement indicators, which are functions of the eigenvalues of a reduced density matrix. The most well-known one is von Neumann entanglement entropy
\begin{equation}\label{eq:vN}
S_{1}=-\tr\left(\rho_{A}\ln\rho_{A}\right)=-\sum_{i=1}^{m}\lambda_{i}\ln\lambda_{i},~~~~~~S_{1}\in\left[0,\ln{m}\right],
\end{equation}
that has been studied under different generic state models in~\cite{Page93,Foong94,Ruiz95,VPO16,Wei17,Wei20,HWC21,SK19,Wei20b,Wei20c,BHK21,BHKRV21,HW22} among other references. The von Neumann entropy monotonically increases from the separable state
\begin{equation}\label{eq:s}
\lambda_{1}=1,~~\lambda_{2}=\dots=\lambda_{m}=0
\end{equation}
when $S_{1}=0$ to the maximally-entangled state
\begin{equation}\label{eq:e}
\lambda_{1}=\lambda_{2}=\dots=\lambda_{m}=\frac{1}{m}
\end{equation}
when $S_{1}=\ln{m}$. Other major entanglement entropies include R\'{e}nyi entropy, quantum purity, and Tsallis entropy, which all satisfy the above monotonicity property~\cite{BZ17}.

The present work focuses on another key entanglement indicator known as capacity of entanglement~\cite{Yao10}
\begin{equation}\label{eq:C}
C=\tr\left(\rho_{A}\ln^{2}\rho_{A}\right)-\tr^{2}\left(\rho_{A}\ln\rho_{A}\right)=S_2-S_{1}^{2},~~~~~~C\in\left[0,C_{\rm{max}}\right],
\end{equation}
where
\begin{equation}\label{eq:S}
S_{k}=(-1)^{k}\sum_{i=1}^{m}\lambda_{i}\ln^{k}\lambda_{i},~~~~k=1,2,\dots,
\end{equation}
defines a family of linear spectral statistics indexed by $k$ with $S_1$ being the von Neumann entropy~(\ref{eq:vN}). Unlike entropies, the capacity vanishes $C=0$ in both separable~(\ref{eq:s}) and maximally-entangled~(\ref{eq:e}) states, whereas it attains the maximum $C=C_{\rm{max}}$ in some partially entangled state\footnote{It is straightforward to verify that, for a given $m$, the value of $C_{\rm{max}}$ can be obtained numerically as the maximum of the function $\left(1-x\right)x\left(\ln(m-1)-\ln\left(1-x\right)+\ln x\right)^2$ in the interval $x\in(0,1)$.}. In the time evolution of quantum systems, the ability to detect the presence of entanglement at earlier times than entropies could capture is a distinguishing characteristics of capacity~\cite{Nandy21}. By the definition~(\ref{eq:C}), the average of capacity
\begin{eqnarray}
\mathbb{E}\!\left[C\right]&=&\mathbb{E}\!\left[S_2\right]-\mathbb{E}\!\left[S_{1}^{2}\right] \label{eq:aC} \\
&=& \mathbb{E}\!\left[\tr\left(\rho_{A}K^{2}\right)\right]-\mathbb{E}\!\left[\tr^{2}\left(\rho_{A}K\right)\right] \label{eq:vK}
\end{eqnarray}
can be also understood as a measure of the fluctuation of the modular Hamiltonian~\cite{Okuyama21}
\begin{equation}\label{eq}
K=-\ln\rho_{A}.
\end{equation}
Therefore, average capacity contains information on the width of the entanglement spectrum, which is otherwise unavailable by inspecting average entropies.

We now present the main results of this work on the exact and asymptotic average capacity under the Hilbert-Schmidt and Bures-Hall ensembles.
\begin{prop}\label{p1}
For a bipartite system of dimensions $m$ and $n$ with the parameter $\alpha$ as defined in~(\ref{eq:a}), the average value of entanglement capacity~(\ref{eq:C}) under the Hilbert-Schmidt ensemble~(\ref{eq:HS}) is given by
\begin{equation}\label{eq:CHS}
\fl\mathbb{E}_{\rm{HS}}\!\left[C\right]=\Psi_{0,\alpha}+a_{0}\psi_{1}(m+\alpha+1)+a_{1}\left(\psi_{0}(m+\alpha+1)-\psi_{0}(\alpha+1)\right)+a_{2},
\end{equation}
where the coefficients $a_0$, $a_1$, and $a_2$ are given by
\begin{eqnarray}
a_0 &=& \frac{(m-1)(m+\alpha-1)}{m(m+\alpha)+1} \\
a_1 &=& \frac{\alpha(2m+\alpha-1)}{m(m+\alpha)} \\
a_2 &=& -\frac{m\left(7m^2+6\alpha m-4m-2\alpha+5\right)}{4(m+\alpha)\left(m^2+\alpha m+1\right)}-1
\end{eqnarray}
and $\Psi_{0,\alpha}$ is a special case of the summation
\begin{equation}\label{eq:sum}
\Psi_{a,b}=\frac{2(m+a)!}{(m+b)!}\sum_{k=1}^{m+a}\frac{(m+b-k)!}{(m+a-k)!}\frac{1}{k^2},~~~~~~a,b\in\mathbb{R}.
\end{equation}
Here, $\psi_{0}(x)=\dd\ln\Gamma(x)/\dd x$ and $\psi_{1}(x)=\dd^{2}\ln\Gamma(x)/\dd x^{2}$ denote respectively the digamma function and trigamma function~\cite{Brychkov08}.
\end{prop}

The proof of proposition~\ref{p1} can be found in section~\ref{sec:CHS}. Note that for integer arguments the digamma function and trigamma function admit respectively the finite summation forms
\begin{equation}\label{eq:pl0}
\psi_{0}(l)=-\gamma+\sum_{k=1}^{l-1}\frac{1}{k}
\end{equation}
and
\begin{equation}
\psi_{1}(l)=\frac{\pi^{2}}{6}-\sum_{k=1}^{l-1}\frac{1}{k^{2}}
\end{equation}
with $\gamma\approx0.5772$ being the Euler's constant. It is also important to point out that the sum~(\ref{eq:sum}) can be written in terms of a hypergeometric function of unit argument as
\begin{equation}\label{eq:4F3}
\Psi_{a,b}=\frac{2(m+a)}{m+b}~\!_{4}F_{3}\left(\begin{array}{c} 1,1,1,1-m-a \\ 2,2,1-m-b
\end{array}\Big|1\Big.\right),
\end{equation}
which in general may not be simplified to a closed-form expression. In the literature, the special cases of~(\ref{eq:CHS}) for small subsystem dimensions $m,n\leq3$ are obtained in~\cite{Boer19}. Another related result is the average formula of annealed capacity $C_A$ derived in~\cite{Okuyama21}. The average of the annealed capacity is defined as~\cite{Okuyama}
\begin{equation}\label{eq:aCA}
\mathbb{E}_{\rm{HS}}\!\left[C_A\right]=\mathbb{E}_{\rm{HS}}\!\left[S_2\right]-\mathbb{E}_{\rm{HS}}^{2}\!\left[S_{1}\right],
\end{equation}
where, comparing to the average capacity~(\ref{eq:aC}), the second-order statistics $\mathbb{E}_{\rm{HS}}\!\left[S_{1}^{2}\right]$ is replaced by a mathematically simpler (squared) first-order statistics $\mathbb{E}_{\rm{HS}}^{2}\!\left[S_{1}\right]$.

Before discussing the results over the Bures-Hall ensemble, we have the following remark.
\begin{remark}
Substituting $m\to m+a$ and $n\to m+b$ in the identity~(\ref{eq:A3}), the sum $\Psi_{a,b}$ in~(\ref{eq:sum}) can be rewritten by the sum
\begin{equation}\label{eq:b1}
2\sum_{k=1}^{m+a}\frac{\psi_{0}(k+b-a)}{k}+\rm{CF},
\end{equation}
where CF denotes the closed-form terms in~(\ref{eq:A3}). The sum in~(\ref{eq:b1}) is known as an unsimplifiable basis~\cite{Wei17,Wei20,HWC21,Wei20b,Wei20c,HW22}, which in general is not summable into a closed-form expression, cf.~(\ref{eq:4F3}). However, in the special cases of given integers $a$ and $b$ with $b\geq a$, the sum~(\ref{eq:b1}) permits closed-form evaluation as a result of the identity~(\ref{eq:A4}). This corresponds to the case of a fixed dimension difference $\alpha=n-m$, where the average capacity~(\ref{eq:CHS}) admits more explicit expressions. The cases $\alpha=0,1,2$ are provided respectively in below as examples
\begin{eqnarray}
\fl\mathbb{E}_{\rm{HS}}\!\left[C\right] &=& -\frac{(m+1)^2}{m^2+1}\psi_{1}(m+1)-\frac{11m^2-4m+9}{4\left(m^2+1\right)}+\frac{\pi^2}{3} \label{eq:a0}\\
\fl\mathbb{E}_{\rm{HS}}\!\left[C\right] &=& -\frac{(m+1)(m+2)}{m^2+m+1}\psi_{1}(m+2)-\frac{11m^2+7m+12}{4\left(m^2+m+1\right)}+\frac{\pi^2}{3} \\
\fl\mathbb{E}_{\rm{HS}}\!\left[C\right] &=& -\frac{m+3}{m+1}\psi_{1}(m+3)+\frac{2(\psi_{0}(m+3)-\psi_{0}(3))}{m(m+1)(m+2)}-\frac{11m^2+29m+28}{4(m+1)(m+2)}+\frac{\pi^2}{3}.\label{eq:a2}
\end{eqnarray}
\end{remark}
We also note that the choice of the sum~(\ref{eq:sum}) over the one in~(\ref{eq:b1}) facilitates the study of the asymptotic capacity as discussed in section~\ref{sec:lim}.

For the Bures-Hall ensemble, the corresponding result is given by the following proposition.
\begin{prop}\label{p2}
For a bipartite system of dimensions $m$ and $n$ with the parameter $\beta$ as defined in~(\ref{eq:b}), the average value of entanglement capacity~(\ref{eq:C}) under the Bures-Hall ensemble~(\ref{eq:BH}) is
\begin{equation}\label{eq:CBH}
\fl\mathbb{E}_{\rm{BH}}\!\left[C\right]=\Psi_{0,\beta}+\Psi_{2\beta,\beta}+b_{0}\psi_{1}(m+\beta+1)+b_{1}\left(\psi_{0}(m+\beta+1)-\psi_{0}(\beta+1)\right)+b_{2},
\end{equation}
where the coefficients $b_0$, $b_1$, and $b_2$ are given by
\begin{eqnarray}
b_0 &=& \frac{2(m-1)(m+\beta)(m+2\beta)}{(2m+2\beta+1)\left(m^2+2\beta m+m+2\right)}+1 \\
b_1 &=& \frac{2\beta^2}{m(m+2\beta+1)} \\
b_2 &=& -\frac{\pi^2}{2}-1
\end{eqnarray}
and $\Psi_{a,b}$ is defined in~(\ref{eq:sum}).
\end{prop}

The proof of proposition~\ref{p2} is in section~\ref{sec:CBH}. For half-integer arguments we have the finite sum representations for the polygamma functions in~(\ref{eq:CBH}) as
\begin{equation}
\psi_{0}\left(l+\frac{1}{2}\right)=-\gamma-2\ln2+2\sum_{k=0}^{l-1}\frac{1}{2k+1}
\end{equation}
\begin{equation}
\psi_{1}\left(l+\frac{1}{2}\right)=\frac{\pi^{2}}{2}-3\sum_{k=1}^{l-1}\frac{1}{k^{2}}-4\sum_{k=l}^{2l-1}\frac{1}{k^{2}}.
\end{equation}
We also note that unlike the cases~(\ref{eq:a0})-(\ref{eq:a2}), here the result~(\ref{eq:CBH}) may not be reduced to a closed-form expression for a fixed $\beta$. This is due to the half-integer nature of $\beta$, in which the sum in~(\ref{eq:b1}) does not lead to a closed-form expression.

Based on the exact capacity formulas~(\ref{eq:CHS}) and~(\ref{eq:CBH}), the corresponding limiting behavior for large dimensional subsystems $m$ and $n$ can also be obtained. The results are summarized in the following corollary.
\begin{coro}\label{c}
For a bipartite system of dimensions $m$ and $n$ in the asymptotic regime
\begin{equation}\label{eq:lim}
m\to\infty,~~~~n\to\infty,~~~~\textrm{with a fixed}~n-m,
\end{equation}
the average entanglement capacity under the Hilbert-Schmidt ensemble~(\ref{eq:CHS}) and the Bures-Hall ensemble~(\ref{eq:CBH}) approach to the limit
\begin{equation}\label{eq:CHSa}
\mathbb{E}_{\rm{HS}}\!\left[C\right]\longrightarrow\frac{\pi^2}{3}-\frac{11}{4}
\end{equation}
and the limit
\begin{equation}\label{eq:CBHa}
\mathbb{E}_{\rm{BH}}\!\left[C\right]\longrightarrow\frac{\pi^2}{6}-1,
\end{equation}
respectively.
\end{coro}

The proof of corollary~\ref{c} is provided in section~\ref{sec:lim}. Note that the limiting values~(\ref{eq:CHSa}) and~(\ref{eq:CBHa}) are independent of $\alpha$ and $\beta$ as defined in~(\ref{eq:a}) and~(\ref{eq:b}), respectively. It is also worth mentioning that for the average annealed capacity~(\ref{eq:aCA}) under the Hilbert-Schmidt ensemble, the limiting behavior derived in~\cite{Okuyama21} turns out the same as~(\ref{eq:CHSa}), i.e.,
\begin{equation}\label{eq:CAHSa}
\mathbb{E}_{\rm{HS}}\!\left[C_A\right]\longrightarrow\frac{\pi^2}{3}-\frac{11}{4}.
\end{equation}
This is because their difference
\begin{equation}\label{eq}
\mathbb{E}_{\rm{HS}}\!\left[C_A\right]-\mathbb{E}_{\rm{HS}}\!\left[C\right]=\mathbb{E}_{\rm{HS}}\!\left[S_{1}^{2}\right]-\mathbb{E}_{\rm{HS}}^{2}\!\left[S_{1}\right]=\mathbb{V}_{\rm{HS}}\!\left[S_1\right],
\end{equation}
which is the variance of von Neumann entropy~(\ref{eq:vHS}), vanishes in the regime~(\ref{eq:lim}) as a result of the limiting behavior of polygamma functions~(\ref{eq:psi0a})-(\ref{eq:psi1a}). For the same reason, since the variance $\mathbb{V}_{\rm{BH}}\!\left[S_1\right]$ under the Bures-Hall ensemble as shown in~(\ref{eq:vBH}) also vanishes in the limit~(\ref{eq:lim}), the resulting average annealed capacity converges to the corresponding asymptotic value~(\ref{eq:CBHa}), i.e.,
\begin{equation}\label{eq:CABHa}
\mathbb{E}_{\rm{BH}}\!\left[C_A\right]\longrightarrow\frac{\pi^2}{6}-1.
\end{equation}
Therefore, the average capacity and average annealed capacity under each considered ensemble are described by the same universal limit despite having distinct finite-size formulas. We also point out that, in the same asymptotic regime~(\ref{eq:lim}), the average capacity (per dimension) under the fermionic Gaussian ensemble of equal dimension subsystems $m=n$ has been derived as~\cite{BNP21}
\begin{equation}\label{eq:CfGa}
\mathbb{E}_{\rm{FG}}\!\left[C\right]\longrightarrow\frac{\pi^2}{8}-1,
\end{equation}
which is conjectured~\cite{HW22} to hold true for arbitrary subsystem dimensions $m\leq n$. The limiting value~(\ref{eq:CfGa}) is argued to serve as a phase transition indicator between integrable and chaotic systems~\cite{BNP21}. It would be interesting to see if a similar argument can be stated to interpret the obtained limits~(\ref{eq:CHSa}) and~(\ref{eq:CBHa}).

\begin{figure}[t!]
\centering
\includegraphics[width=0.87\linewidth]{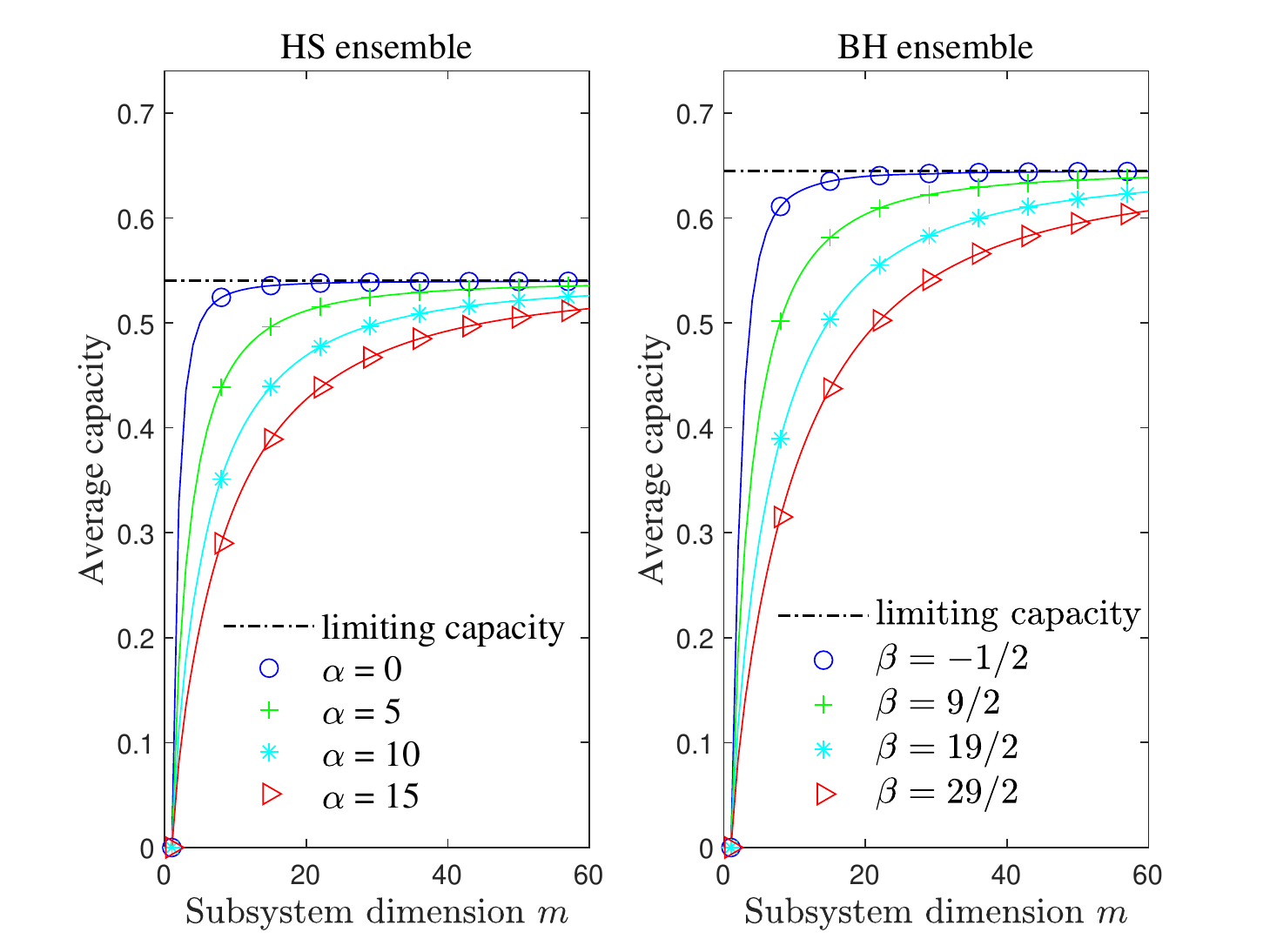}
\caption{Average capacity under HS and BH ensembles: analytical results versus simulations. The solid lines are drawn by the exact capacity formulas~(\ref{eq:CHS}) and~(\ref{eq:CBH}), while the dash-dot horizontal lines represent the limiting values of capacity~(\ref{eq:CHSa}) and~(\ref{eq:CBHa}). The corresponding scatters, as denoted by the symbols of circle, plus sign, asterisk, and triangle, are obtained from numerical simulations.}
\label{fig:p1}
\end{figure}

As a direct consequence of corollary~\ref{c} and the known asymptotic behavior of average entropy in the limit~(\ref{eq:lim}), cf.~(\ref{eq:mHS}),~(\ref{eq:mBH}), and~(\ref{eq:psi0a})-(\ref{eq:psi1a}),
\begin{eqnarray}
\mathbb{E}_{\rm{HS}}\!\left[S_1\right]&=&\ln m+o\left(\frac{1}{m}\right) \label{eq:mHSa} \\
\mathbb{E}_{\rm{BH}}\!\left[S_1\right]&=&\ln m+o\left(\frac{1}{m}\right), \label{eq:mBHa}
\end{eqnarray}
the relative rate of growth between the average capacity and the average entropy, relevant to quantum field theory~\cite{Boer19}, can now be rigorously found as
\begin{equation}\label{eq:CoE}
\frac{\mathbb{E}_{\rm{HS}}\!\left[C\right]}{\mathbb{E}_{\rm{HS}}\!\left[S_1\right]} \stackrel{(\ref{eq:lim})}{\longrightarrow} 0
\end{equation}
\begin{equation}
\frac{\mathbb{E}_{\rm{BH}}\!\left[C\right]}{\mathbb{E}_{\rm{BH}}\!\left[S_1\right]} \stackrel{(\ref{eq:lim})}{\longrightarrow} 0,
\end{equation}
where the result~(\ref{eq:CoE}) under the Hilbert-Schmidt ensemble has been suggested in~\cite{Boer19}.

To illustrate the obtained main results, we plot in figure~\ref{fig:p1} the formulas~(\ref{eq:CHS}),~(\ref{eq:CBH}), and~(\ref{eq:CHSa})-(\ref{eq:CBHa}) as a function of subsystem dimension $m$ when fixing $\alpha$ and $\beta$ as compared with numerical simulations. The left-hand side and right-hand side subfigures correspond to the cases of the Hilbert-Schmidt and Bures-Hall ensembles, respectively. We observe that as the dimension differences $\alpha$ and $\beta$ increase, the corresponding average capacity~(\ref{eq:CHS}) and~(\ref{eq:CBH}) approach to the respective limiting values~(\ref{eq:CHSa}) and~(\ref{eq:CBHa}) more slowly. In other words, the finite-size capacity formulas are more useful when $\alpha$ and $\beta$ are large, and otherwise the asymptotic capacity values serve as good-enough approximations. By comparing the two subfigures, it is also observed that the average capacity under the Bures-Hall ensemble attains a larger value for a given subsystem dimensions. By the variance of the modular Hamiltonian interpretation~(\ref{eq:vK}) of the capacity, this observation implies that the width of the spectrum of Bures-Hall ensemble is on average wider than that of the Hilbert-Schmidt ensemble. The numerical simulations as represented by the scatters, each obtained over $10^6$ realizations of random density matrices, match quite well with the analytical results as expected.

\section{Derivation of average capacity formulas}\label{sec:proof}
In this section, we provide detailed derivations of the claimed results in the previous section. We first present a lemma in section~\ref{sec:r} that relates the average capacity computation to averages over some induced ensembles. The resulting average capacity expressions of the Hilbert-Schmidt ensemble in proposition~\ref{p1} and the Bures-Hall ensemble in proposition~\ref{p2} are derived in section~\ref{sec:CHS} and section~\ref{sec:CBH}, respectively. The limiting values of capacity in corollary~\ref{c} are computed in section~\ref{sec:lim}. For convenience, we also summarize the main contribution of this work in table~\ref{t:1} below.

\begin{longtable}[h!]{lcc}
\caption{Main results} \\
\ns\hline\hline
\bs average capacity & exact & asymptotic \\
\bs\bs under Hilbert-Schmidt ensemble & eq.(\ref{eq:CHS}) & {\large $\frac{\pi^2}{3}-\frac{11}{4}$} \\
\bs\bs under Bures-Hall ensemble & eq.(\ref{eq:CBH}) & {\large $\frac{\pi^2}{6}-1$} \\
\bs\hline\hline
\label{t:1}
\end{longtable}

\subsection{Average capacity relation}\label{sec:r}
To compute the average capacity~(\ref{eq:aC}) is essentially to compute the average $\mathbb{E}\!\left[S_2\right]$ as the second moment $\mathbb{E}\!\left[S_{1}^{2}\right]$ is known under the Hilbert-Schmidt ensemble~(\ref{eq:mHS})-(\ref{eq:vHS}) and the Bures-Hall ensemble~(\ref{eq:mBH})-(\ref{eq:vBH}). As shown in the following lemma, the averages of $\mathbb{E}\!\left[S_2\right]$ can be converted to these of the induced entropies
\begin{equation}\label{eq:T}
T_{k}=\sum_{i=1}^{m}x_{i}\ln^{k}x_{i},~~~~k=1,2,\dots,
\end{equation}
over certain induced ensembles, whose density functions of an arbitrary eigenvalue are available.

\begin{lemma}\label{l}
The averages of $S_2$ defined in~(\ref{eq:S}) over the Hilbert-Schmidt ensemble~(\ref{eq:HS}) and the Bures-Hall ensemble~(\ref{eq:BH}) are related to the averages of $T_{2}$ in~(\ref{eq:T}) over the respective induced ensembles~(\ref{eq:iHS}) and~(\ref{eq:iBH}) as
\begin{equation}\label{eq:ST}
\mathbb{E}\!\left[S_2\right]=\frac{1}{d}\mathbb{E}\!\left[T_2\right]+2\psi_{0}(d+1)\mathbb{E}\!\left[S_1\right]-\psi_{0}^{2}(d+1)-\psi_{1}(d+1),
\end{equation}
where
\begin{equation}\label{eq:d}
\label{cases}
d=\cases{mn, & Hilbert-Schmidt ensemble \\ \frac{1}{2}m(2n-m-1), & Bures-Hall ensemble \\}
\end{equation}
\end{lemma}
\textit{Proof.}~~The starting point of the proof is the relation~\cite{Page93,Wei17,Wei20b,Wei20c} between the ensembles~(\ref{eq:HS}),~(\ref{eq:BH}) and the corresponding induced ones~(\ref{eq:iHS}),~(\ref{eq:iBH})
\begin{equation}\label{eq:fhg}
f(\bm{\lambda})h_{d}(r)\dd\bm{\lambda}\dd r=g(\bm{x})\dd\bm{x}
\end{equation}
under the change of variables
\begin{equation}\label{eq:lx}
\lambda_{i}=\frac{x_{i}}{r},~~~~i=1,\dots,m,
\end{equation}
where
\begin{equation}\label{eq:h}
h_{d}(r)=\frac{1}{\Gamma(d)}\e^{-r}r^{d-1},~~~~~~r\in[0,\infty),
\end{equation}
is the density of the trace
\begin{equation}\label{eq:tr}
r=\sum_{i=1}^{m}x_{i}
\end{equation}
with the parameter $d$ given by~(\ref{eq:d}) as obtained in~\cite{Wei17,Wei20c}. The relation~(\ref{eq:fhg}) implies that $r$ is independent of $\bm{\lambda}$, and is hence independent of $S_{k}$ for each $k$. To exploit this fact, one rewrites $S_{2}$ by using the change of variables~(\ref{eq:lx}) as, cf.~\cite{Wei17,Wei20c},
\begin{equation}\label{eq:S2T2}
S_{2}=\frac{1}{r}T_{2}+2S_{1}\ln r-\ln^{2}r.
\end{equation}
Consequently, by multiplying with the density~(\ref{eq:h}) of an appropriate choice of the parameter $d$ so as to utilize the relation~(\ref{eq:fhg}), we have
\begin{eqnarray}
\mathbb{E}\!\left[S_2\right] &=& \int_{\bm{\lambda}}\left(\frac{1}{r}T_{2}+2S_{1}\ln r-\ln^{2}r\right)f(\bm{\lambda})\dd\bm{\lambda}\int_{r}h_{d+1}(r)\dd r \\
&=& \frac{1}{d}\mathbb{E}\!\left[T_2\right]+\frac{2}{d}\mathbb{E}\!\left[r\ln r\right]\mathbb{E}\!\left[S_{1}\right]-\frac{1}{d}\mathbb{E}\!\left[r\ln^{2}r\right],
\end{eqnarray}
where the expectation of $S_{1}\ln r$ factorizes due to the independence. The averages over the density of trace~(\ref{eq:h}) are easily computed, see for example~\cite{Wei17,Wei20c}, as
\begin{eqnarray}
\mathbb{E}\!\left[r\ln r\right] &=& d\psi_{0}(d+1)\\
\mathbb{E}\!\left[r\ln^{2}r\right] &=& d\left(\psi_{0}^{2}(d+1)+\psi_{1}(d+1)\right).
\end{eqnarray}
This completes the proof of lemma~\ref{l}.

\subsection{Average capacity over Hilbert-Schmidt ensemble}\label{sec:CHS}
We now prove proposition~\ref{p1}. To compute the average capacity over the Hilbert-Schmidt ensemble
\begin{equation}\label{eq:CHS12}
\mathbb{E}_{\rm{HS}}\!\left[C\right]=\mathbb{E}_{\rm{HS}}\!\left[S_2\right]-\mathbb{E}_{\rm{HS}}\!\left[S_{1}^{2}\right],
\end{equation}
we first note that its second term
\begin{equation}\label{eq}
\mathbb{E}_{\rm{HS}}\!\left[S_{1}^{2}\right]=\mathbb{V}_{\rm{HS}}\!\left[S_1\right]+\mathbb{E}_{\rm{HS}}^{2}\!\left[S_1\right],
\end{equation}
is known in the literature, where
\begin{equation}\label{eq:mHS}
\mathbb{E}_{\rm{HS}}\!\left[S_1\right]=\psi_{0}(mn+1)-\psi_{0}(n)-\frac{m+1}{2n}
\end{equation}
and
\begin{equation}\label{eq:vHS}
\mathbb{V}_{\rm{HS}}\!\left[S_1\right]=-\psi_{1}\left(mn+1\right)+\frac{m+n}{mn+1}\psi_{1}\left(n\right)-\frac{(m+1)(m+2n+1)}{4n^{2}(mn+1)}
\end{equation}
have been obtained in~\cite{Page93,Ruiz95} and~\cite{VPO16,Wei17}, respectively. By lemma~\ref{l}, computing the first term $\mathbb{E}_{\rm{HS}}\!\left[S_2\right]$ in~(\ref{eq:CHS12}) boils down to computing $\mathbb{E}_{\rm{HS}}\!\left[T_2\right]$ as the average $\mathbb{E}_{\rm{HS}}\!\left[S_1\right]$ in~(\ref{eq:ST}) is available~(\ref{eq:mHS}). To compute $\mathbb{E}_{\rm{HS}}\!\left[T_2\right]$ requires
an arbitrary eigenvalue density $p_{\rm{HS}}(x)$ of the so-defined induced\footnote{This new ensemble is induced by the change of measures~(\ref{eq:fhg}).} Hilbert-Schmidt ensemble
\begin{equation}\label{eq:iHS}
g_{\rm{HS}}\left(\bm{x}\right)\propto
\prod_{1\leq i<j\leq m}\left(x_{i}-x_{j}\right)^{2}\prod_{i=1}^{m}x_{i}^{\alpha}\e^{-x_{i}},~~~~~~x_{i}\in[0,\infty),
\end{equation}
given by~\cite{Wei17}
\begin{eqnarray}\label{eq:iHS1p}
p_{\rm{HS}}(x)=\frac{(m-1)!}{(m+\alpha-1)!}x^{\alpha}\e^{-x}\left(\left(L_{m-1}^{(\alpha+1)}(x)\right)^{2}-L_{m-2}^{(\alpha+1)}(x)L_{m}^{(\alpha+1)}(x)\right),
\end{eqnarray}
where we recall that the parameter $\alpha$ denotes the dimension difference~(\ref{eq:a}). In~(\ref{eq:iHS1p}), the function $L_{k}^{(\alpha)}(x)$ is the Laguerre orthogonal polynomial of degree $k$,
\begin{equation}
L_{k}^{(\alpha)}(x)=\sum_{i=0}^{k}(-1)^{i}{\alpha+k\choose k-i}\frac{x^i}{i!}
\end{equation}
that satisfies the orthogonality relation~\cite{Mehta,Forrester}
\begin{equation}\label{eq:oc}
\int_{0}^{\infty}\!\!x^{\alpha}\e^{-x}L_{k}^{(\alpha)}(x)L_{l}^{(\alpha)}(x)\dd{x}=\frac{(\alpha+k)!}{k!}\delta_{kl}
\end{equation}
with $\delta_{kl}$ being the Kronecker delta function.

We now have
\begin{eqnarray}
\mathbb{E}_{\rm{HS}}\!\left[T_2\right] &=& m\int_{0}^{\infty}\!\!x\ln^{2}x~p_{\rm{HS}}(x)\dd x \\
&=&\mathcal{A}_{m-1,m-1}-\mathcal{A}_{m-2,m}, \label{eq:iHS1}
\end{eqnarray}
where $\mathcal{A}_{s,t}$ denotes the integral
\begin{equation}\label{eq:Ast}
\mathcal{A}_{s,t}=\frac{m!}{(m+\alpha-1)!}\int_{0}^{\infty}\!\!x^{\alpha+1}\e^{-x}\ln^{2}{x}~L_{s}^{(\alpha+1)}(x)L_{t}^{(\alpha+1)}(x)\dd{x}.
\end{equation}
The above integral~(\ref{eq:Ast}) can be evaluated by using the identity
\begin{eqnarray}\label{eq:SI2}
&&\int_{0}^{\infty}\!\!x^{q}\e^{-x}\ln^{2}{x}~L_{s}^{(a)}(x)L_{t}^{(b)}(x)\dd{x} \nonumber \\
&=&(-1)^{s+t}\sum_{k=0}^{\min(s,t)}{q-a\choose s-k}{q-b\choose t-k}\frac{\Gamma(q+1+k)}{k!}\left(\Omega_{0}^{2}+\Omega_{1}\right),
\end{eqnarray}
where we denote
\begin{eqnarray}
\Omega_{j}&=&\psi_{j}(q+1+k)+\psi_{j}(q-a+1)+\psi_{j}(q-b+1) \nonumber \\
&&-\psi_{j}(q-a-s+1+k)-\psi_{j}(q-b-t+1+k).
\end{eqnarray}
The identity~(\ref{eq:SI2}) is obtained by taking twice derivatives with respect to $q$ of an integral identity of Schr{\"{o}}dinger~\cite{Schrodinger1926}
\begin{equation}\label{eq:Swm}
\fl\int_{0}^{\infty}\!\!x^{q}\e^{-x}L_{s}^{(a)}(x)L_{t}^{(b)}(x)\dd{x}=(-1)^{s+t}\sum_{k=0}^{\min(s,t)}{q-a\choose s-k}{q-b\choose t-k}\frac{\Gamma(q+1+k)}{k!},
\end{equation}
valid for $\Re(q)>-1$, which is a generalization of the orthogonality relation~(\ref{eq:oc}).

With the specialization $a=b=\alpha+1$ and $q=\alpha+1$ in~(\ref{eq:SI2}), after resolving the indeterminacy in the limit $\epsilon\to0$ by using
\begin{eqnarray}
\Gamma(-l+\epsilon)&=&\frac{(-1)^{l}}{l!\epsilon}\left(1+\psi_{0}(l+1)\epsilon+o\left(\epsilon^2\right)\right) \label{eq:pgna1}\\
\psi_{0}(-l+\epsilon)&=&-\frac{1}{\epsilon}\left(1-\psi_{0}(l+1)\epsilon+o\left(\epsilon^2\right)\right) \label{eq:pgna2}\\
\psi_{1}(-l+\epsilon)&=&\frac{1}{\epsilon^2}\left(1+o\left(\epsilon^2\right)\right), \label{eq:pgna3}
\end{eqnarray}
the two integrals in~(\ref{eq:iHS1}) are evaluated into finite summations of the forms
\begin{equation}\label{eq:A1s}
\sum_{k=1}^{n}k^{c}\psi_{j}^{b}(k+a),~~~~j=0,1,
\end{equation}
and
\begin{equation}\label{eq:A2s}
\sum_{k=1}^{m}\frac{(n-k)!}{(m-k)!}\frac{1}{k^j},~~~~j=0,1,2,
\end{equation}
where $a,b,c$ are non-negative integers. By using the relevant summation identities of the type~(\ref{eq:A1s}) listed in~\cite{Wei20} and the type~(\ref{eq:A2s}) in~(\ref{eq:A1})-(\ref{eq:A2}), we obtain
\begin{eqnarray}\label{eq:TCHS1}
\mathcal{A}_{m-1,m-1}&=&\frac{2m!m}{(n-1)!}\sum_{k=1}^{m}\frac{(n-k)!}{(m-k)!}\frac{1}{k^2}+mn\left(\psi_{0}^{2}(n+1)+\psi_1(n+1)\right)\nonumber \\
&&+2n\left(\psi_{0}(n-m+1)-\psi_{0}(n+1)\right)
\end{eqnarray}
and
\begin{eqnarray}\label{eq:TCHS2}
\mathcal{A}_{m-2,m}&=&\left(n^2+n-m^2+m\right)\psi_{0}(n-m+1)-n(n+1)\psi_{0}(n+1) \nonumber \\
&&+\frac{1}{2}m(2n+3m-1).
\end{eqnarray}
Finally, substituting the results~(\ref{eq:mHS})-(\ref{eq:vHS}),~(\ref{eq:ST})-(\ref{eq:d}), and~(\ref{eq:TCHS1})-(\ref{eq:TCHS2}) in~(\ref{eq:CHS12}) completes the proof of proposition~\ref{p1}.

\subsection{Average capacity over Bures-Hall ensemble}\label{sec:CBH}
Here, we prove proposition~\ref{p2}. The second term
\begin{equation}\label{eq}
\mathbb{E}_{\rm{BH}}\!\left[S_{1}^{2}\right]=\mathbb{V}_{\rm{BH}}\!\left[S_1\right]+\mathbb{E}_{\rm{BH}}^{2}\!\left[S_1\right]
\end{equation}
of the average capacity over the Bures-Hall ensemble
\begin{equation}\label{eq:CBH12}
\mathbb{E}_{\rm{BH}}\!\left[C\right]=\mathbb{E}_{\rm{BH}}\!\left[S_2\right]-\mathbb{E}_{\rm{BH}}\!\left[S_{1}^{2}\right]
\end{equation}
has been recently computed~\cite{SK19,Wei20b,Wei20c} as shown in~(\ref{eq:mBH}) and~(\ref{eq:vBH}) below.
\begin{equation}\label{eq:mBH}
\mathbb{E}_{\rm{BH}}\!\left[S_1\right]=\psi_{0}\left(mn-\frac{m^2}{2}+1\right)-\psi_{0}\left(n+\frac{1}{2}\right)
\end{equation}
\begin{equation}\label{eq:vBH}
\mathbb{V}_{\rm{BH}}\!\left[S_1\right]=-\psi_{1}\left(mn-\frac{m^2}{2}+1\right)+\frac{2n(2n+m)-m^{2}+1}{2n(2mn-m^2+2)}\psi_{1}\left(n+\frac{1}{2}\right)
\end{equation}
By lemma~\ref{l}, now the task is to calculate the average $\mathbb{E}_{\rm{BH}}\!\left[T_2\right]$ over the ensemble $g_{\rm{BH}}\left(\bm{x}\right)$ induced from the change of measures~(\ref{eq:fhg}), which is given by~\cite{SK19,Wei20b,Wei20c}
\begin{equation}\label{eq:iBH}
g_{\rm{BH}}\left(\bm{x}\right)\propto
\prod_{1\leq i<j\leq m}\frac{\left(x_{i}-x_{j}\right)^{2}}{x_{i}+x_{j}}\prod_{i=1}^{m}x_{i}^{\beta}\e^{-x_{i}},~~~~~~x_{i}\in[0,\infty).
\end{equation}
It was recently discovered in~\cite{FK16} that the correlation functions of the induced ensemble~(\ref{eq:iBH}) can be written in terms of these of the Cauchy-Laguerre biorthogonal ensemble~\cite{BGS14}. In particular, the needed density of an arbitrary eigenvalue is~\cite{FK16}
\begin{equation}\label{eq:iBH1p}
p_{\rm{BH}}(x)=\frac{1}{2m}\left(K_{01}(x,x)+K_{10}(x,x)\right),
\end{equation}
where the kernels $K_{01}(x,y)$ and $K_{10}(x,y)$ admit the integral representations~\cite{BGS14}
\begin{eqnarray}
K_{01}(x,y)&=&x^{2\beta+1}\int_{0}^{1}t^{2\beta+1}H_{\beta}(ty)G_{\beta+1}(tx)\dd t \label{eq:K01} \\
K_{10}(x,y)&=&y^{2\beta+1}\int_{0}^{1}t^{2\beta+1}H_{\beta+1}(tx)G_{\beta}(ty)\dd t. \label{eq:K10}
\end{eqnarray}
Here, the functions
\begin{eqnarray}
H_{q}(x)&=&G_{2,3}^{1,1}\left(\!\begin{array}{c}-m-2\beta-1;m\\0;-q,-2\beta-1\end{array}\Big|x\Big.\right) \label{eq:Hq} \\
G_{q}(x)&=&G_{2,3}^{2,1}\left(\!\begin{array}{c}-m-2\beta-1;m\\0,-q;-2\beta-1\end{array}\Big|x\Big.\right)
\end{eqnarray}
are the Meijer G-function, which in general is defined by the contour integral~\cite{PBM86}
\begin{eqnarray}\label{eq:MG}
&&G_{p,q}^{m,n}\left(\begin{array}{c} a_{1},\ldots,a_{n}; a_{n+1},\ldots,a_{p} \\ b_{1},\ldots,b_{m}; b_{m+1},\ldots,b_{q} \end{array}\Big|x\Big.\right)\nonumber\\
&&=\frac{1}{2\pi\imath}\int_{\mathcal{L}}{\frac{\prod_{j=1}^m\Gamma\left(b_j+s\right)\prod_{j=1}^n\Gamma\left(1-a_j-s\right)x^{-s}}{\prod_{j=n+1}^p \Gamma\left(a_{j}+s\right)\prod_{j=m+1}^q\Gamma\left(1-b_j-s\right)}}\dd s,
\end{eqnarray}
with the poles of $\Gamma\left(1-a_j-s\right)$ being separated by the contour $\mathcal{L}$ from the poles of $\Gamma\left(b_j+s\right)$. By using the fact that the Meijer G-function in~(\ref{eq:Hq}) can be written as a terminating hypergeometric function~\cite{BGS14,FK16}
\begin{eqnarray}
\fl G_{2,3}^{1,1}\left(\!\begin{array}{c}-m-2\beta-1;m\\0;-q,-2\beta-1\end{array}\Big|x\Big.\right) &=& \frac{\Gamma(m+2\beta+2)}{\Gamma(m)\Gamma(q+1)\Gamma(2\beta+2)}~\!_{2}F_{2}\left(\begin{array}{c} m+2\beta+2, 1-m \\ q+1, 2\beta+2
\end{array}\Big|x\Big.\right)\nonumber \\
&=& \sum_{k=0}^{m-1}\frac{\Gamma(m+2\beta+2+k)(-x)^{k}}{\Gamma(q+1+k)\Gamma(2\beta+2+k)\Gamma(m-k)k!}
\end{eqnarray}
and the following integral of the Meijer G-function~\cite{PBM86}
\begin{eqnarray}
&&\int_{0}^{1}\!x^{a-1}G_{p,q}^{m,n}\left(\begin{array}{c} a_{1},\ldots,a_{n}; a_{n+1},\ldots,a_{p} \\ b_{1},\ldots,b_{m}; b_{m+1},\ldots,b_{q} \end{array}\Big|\eta x\Big.\right)\dd x \nonumber\\
&=&G_{p+1,q+1}^{m,n+1}\left(\begin{array}{c}1-a,a_{1},\ldots,a_{n};a_{n+1},\ldots,a_{p}\\b_{1},\ldots,b_{m};b_{m+1},\ldots,b_{q},-a\end{array}\Big|\eta\Big.\right),
\end{eqnarray}
the integrals over $t$ of the kernels~(\ref{eq:K01}) and~(\ref{eq:K10}) can be evaluated first. This leads to
\begin{eqnarray}
K_{01}(x,y) &=& \sum_{k=0}^{m-1}\frac{\Gamma(m+2\beta+2+k)(-y)^{k}}{\Gamma(\beta+1+k)\Gamma(2\beta+2+k)\Gamma(m-k)k!}F_{\beta+1}(x) \label{eq:K01s} \\
K_{10}(x,y) &=& \sum_{k=0}^{m-1}\frac{\Gamma(m+2\beta+2+k)(-x)^{k}}{\Gamma(\beta+2+k)\Gamma(2\beta+2+k)\Gamma(m-k)k!}F_{\beta}(y), \label{eq:K10s}
\end{eqnarray}
where we denote
\begin{equation}\label{eq}
F_{q}(x)=G_{3,4}^{2,2}\left(\!\begin{array}{c}-k,-m;m+2\beta+1 \\ 2\beta+1,2\beta+1-q; 0, -k-1\end{array}\Big|x\Big.\right).
\end{equation}

With the above results, we now have
\begin{eqnarray}\label{eq}
\fl \mathbb{E}_{\rm{BH}}\!\left[T_2\right] &=& m\int_{0}^{\infty}\!\!x\ln^{2}x~p_{\rm{BH}}(x)\dd x \\
\fl &=& \frac{1}{2}\int_{0}^{\infty}\!\!x\ln^{2}x K_{01}(x,x)\dd x+\frac{1}{2}\int_{0}^{\infty}\!\!x\ln^{2}x K_{10}(x,x)\dd x \\
\fl &=& \frac{1}{2}\left(\frac{\dd}{\dd\gamma^{2}}\int_{0}^{\infty}\!\!x^{\gamma} K_{01}(x,x)\dd x\right)\Big|_{\gamma=1}+\frac{1}{2}\left(\frac{\dd}{\dd\gamma^{2}}\int_{0}^{\infty}\!\!x^{\gamma} K_{10}(x,x)\dd x\right)\Big|_{\gamma=1}, \label{eq:2T2}
\end{eqnarray}
where, upon substituting~(\ref{eq:K01s}) and~(\ref{eq:K10s}) in~(\ref{eq:2T2}), the calculation proceeds by first invoking the Mellin transform of Meijer G-function~\cite{PBM86}
\begin{eqnarray}\label{eq:iMG}
&&\int_{0}^{\infty}x^{s-1}G_{p,q}^{m,n}\left(\begin{array}{c} a_{1},\ldots,a_{n}; a_{n+1},\ldots,a_{p} \\ b_{1},\ldots,b_{m}; b_{m+1},\ldots,b_{q} \end{array}\Big|\eta x\Big.\right)\dd x \nonumber \\
&=&\frac{\eta^{-s}\prod_{j=1}^m\Gamma\left(b_j+s\right)\prod_{j=1}^{n}\Gamma\left(1-a_j-s\right)}{\prod_{j=n+1}^{p}\Gamma\left(a_{j}+s\right)\prod_{j=m+1}^q\Gamma\left(1-b_j-s\right)}
\end{eqnarray}
before taking the required derivatives and setting $\gamma=1$. We insert the resulting expression into~(\ref{eq:ST}) and~(\ref{eq:CBH12}), the average capacity is obtained as
\begin{eqnarray}\label{eq:CBHus}
\fl&&\mathbb{E}_{\rm{BH}}\!\left[C\right]=2\sum_{k=1}^{m}\left(\frac{\psi_{0}(k)+\psi_{0}(2\beta+k)+\psi_{0}(\beta+k)}{m+2\beta+k}-\frac{\psi_{0}(k)}{\beta+k}-\frac{\psi_{0}(k)}{2\beta+k}\right)\nonumber\\
\fl&&-\left(\psi_{0}(\beta+1)-\psi_{0}(\beta+m+1)\right)^2-\left(\psi_{0}(2\beta+1)-\psi_{0}(m+2\beta+1)\right)^2 \nonumber\\
\fl&&+2\left(\psi_{0}(m+\beta+1)+\psi_{0}(m+2\beta+1)\right)\left(\psi_{0}(m+2\beta+1)-\psi_{0}(2m+2\beta+1)\right) \nonumber\\
\fl&&-2(\psi_{0}(\beta+1)+\psi_{0}(2\beta+1)-\psi_{0}(m+\beta+1)-2\psi_{0}(m+2\beta+1)+\psi_{0}(2m+2\beta+1)) \nonumber\\
\fl&&\times\psi_{0}(m+1)+\psi_1(\beta+1)+\psi_1(2\beta+1)-\psi_1(m+2\beta+1)-\psi_1(m+\beta+1) \nonumber\\
\fl&&\times\frac{5m^2+10m\beta+5m+4\beta^2+4\beta+2}{(2m+2\beta+1)\left(m^2+2m\beta+m+2\right)}+\frac{2\beta^{2}\left(\psi_{0}(m+\beta+1)-\psi_{0}(\beta+1)\right)}{m(m+2\beta+1)}.
\end{eqnarray}
In obtaining the above result~(\ref{eq:CBHus}), we have also used some summation identities of the types~(\ref{eq:A1s}) and~(\ref{eq:A2s}) listed in appendix A as well as in~\cite{Wei20} after resolving the indeterminacy by using~(\ref{eq:pgna1})-(\ref{eq:pgna3}).

The remaining task is to represent each of the five summations in~(\ref{eq:CBHus}),
\begin{eqnarray}
&&\sum_{k=1}^{m}\frac{\psi_{0}(k)}{m+2\beta+k} \label{eq:sum1} \\
&&\sum_{k=1}^{m}\frac{\psi_{0}(2\beta+k)}{m+2\beta+k} \label{eq:sum2} \\
&&\sum_{k=1}^{m}\frac{\psi_{0}(\beta+k)}{m+2\beta+k} \label{eq:sum3} \\
&&\sum_{k=1}^{m}\frac{\psi_{0}(k)}{\beta+k} \label{eq:sum4} \\
&&\sum_{k=1}^{m}\frac{\psi_{0}(k)}{2\beta+k} \label{eq:sum5}
\end{eqnarray}
into the summation $\Psi_{a,b}$ in~(\ref{eq:sum}) as reproduced below
\begin{equation}\label{eq:suma}
\Psi_{a,b}=\frac{2(m+a)!}{(m+b)!}\sum_{k=1}^{m+a}\frac{(m+b-k)!}{(m+a-k)!}\frac{1}{k^2}.
\end{equation}
Namely, we will show that the sums~(\ref{eq:sum1})-(\ref{eq:sum5}), despite of no closed-form evaluations, can all be rewritten in the same form~(\ref{eq:suma}). Firstly, the sum~(\ref{eq:sum1}) is processed by using~(\ref{eq:Aab}) before applying~(\ref{eq:A3}) that results in
\begin{eqnarray}\label{eq}
\sum_{k=1}^{m}\frac{\psi_{0}(k)}{m+2\beta+k} &=&-\sum_{k=1}^{m}\frac{\psi_{0}(m+2\beta+k)}{k}+\rm{CF} \\
&=&-\frac{m!}{(2m+2\beta)!}\sum_{k=1}^{m}\frac{(2m+2\beta-k)!}{(m-k)!}\frac{1}{k^2}+\rm{CF} \\
&=&-\frac{1}{2}\Psi_{0,m+2\beta}+\rm{CF}. \label{eq:sum1f}
\end{eqnarray}
As the same in~(\ref{eq:b1}), the closed-form terms are denoted by the shorthand notation CF, which in general is different in each appearance. In the same manner, the sums~(\ref{eq:sum4}) and~(\ref{eq:sum5}) are rewritten respectively as
\begin{equation}\label{eq}
\sum_{k=1}^{m}\frac{\psi_{0}(k)}{\beta+k}=-\frac{1}{2}\Psi_{0,\beta}+\rm{CF} \label{eq:sum2f}
\end{equation}
and
\begin{equation}\label{eq}
\sum_{k=1}^{m}\frac{\psi_{0}(k)}{2\beta+k}=-\frac{1}{2}\Psi_{0,2\beta}+\rm{CF}. \label{eq:sum3f}
\end{equation}
Additional ingredients in processing the sum~(\ref{eq:sum2}) are the fact, cf.~(\ref{eq:pl0}),
\begin{equation}\label{eq:0s}
\psi_{0}(l+n)=\psi_{0}(l)+\sum_{k=0}^{n-1}\frac{1}{l+k}
\end{equation}
and the identity~(\ref{eq:A4}), where the manipulation proceeds as
\begin{eqnarray}
\sum_{k=1}^{m}\frac{\psi_{0}(2\beta+k)}{m+2\beta+k} &=& -\sum_{k=1}^{m}\frac{\psi_{0}(m+2\beta+k)}{2\beta+k}+\rm{CF} \\
&=& -\sum_{k=1}^{m}\frac{\psi_{0}(2\beta+k)+\sum_{i=1}^{m}\frac{1}{2\beta+k+i-1}}{2\beta+k}+\rm{CF} \\
&=& -\sum _{k=1}^m \frac{\psi_{0}(2\beta+k)}{k}+\sum_{k=1}^{m}\frac{\psi_{0}(m+2\beta+k)}{k}+\rm{CF} \\
&=& -\frac{1}{2}\Psi_{0,2\beta}+\frac{1}{2}\Psi_{0,m+2\beta}+\rm{CF}. \label{eq:sum4f}
\end{eqnarray}
Similarly, for the remaining sum~(\ref{eq:sum3}), we have
\begin{eqnarray}
\sum_{k=1}^{m}\frac{\psi_{0}(\beta+k)}{m+2\beta+k} &=& -\sum_{k=1}^{m}\frac{\psi_{0}(m+2\beta+k)}{\beta+k}+\rm{CF} \\
&=& -\sum_{k=1}^{m}\frac{\psi_{0}(2\beta+k)}{\beta+k}+\rm{CF} \\
&=& \sum_{k=1}^{m}\frac{\psi_{0}(\beta+k)}{2\beta+k}+\rm{CF} \\
&=& \sum_{k=1}^{m+2\beta}\frac{\psi_{0}(k-\beta)}{k}-\sum_{k=1}^{2\beta}\frac{\psi_{0}(k-\beta)}{k}+\rm{CF} \\
&=& \frac{(m+2\beta)!}{(m+\beta)!}\sum_{k=1}^{m+2\beta}\frac{(m+\beta-k)!}{(m+2\beta-k)!}\frac{1}{k^2}+\rm{CF} \label{eq:sum5fi} \\
&=& \frac{1}{2}\Psi_{2\beta,\beta}+\rm{CF}, \label{eq:sum5f}
\end{eqnarray}
where we used the closed-form identity~(\ref{eq:Ab}) in arriving at~(\ref{eq:sum5fi}). Finally, inserting~(\ref{eq:sum1f})-(\ref{eq:sum3f}),~(\ref{eq:sum4f}), and~(\ref{eq:sum5f}) into the capacity expression~(\ref{eq:CBHus}) leads to substantial cancellations, where the remaining terms are given by~(\ref{eq:CBH}). This completes the proof of proposition~\ref{p2}.

\subsection{Asymptotic capacity}\label{sec:lim}
Computing the limiting capacity~(\ref{eq:CHSa}) and~(\ref{eq:CBHa}) in corollary~\ref{c} is a rather straightforward task, which requires computing the limits in the regime~(\ref{eq:lim}) of the respective exact capacity~(\ref{eq:CHS}) and~(\ref{eq:CBH}). One first computes the limit of the summation~(\ref{eq:sum}) as
\begin{equation}
\Psi_{a,b} \stackrel{(\ref{eq:lim})}{\longrightarrow} 2\sum_{k=1}^{m}\frac{1}{k^2}=2\psi_{1}(1)=\frac{\pi^2}{3},
\end{equation}
which is valid for any finite $a$ and $b$. We also need the limiting behavior of polygamma functions~\cite{Brychkov08}
\begin{eqnarray}
\psi_0(x)&=&\ln x-\frac{1}{2 x}-\sum_{l=1}^{\infty}\frac{B_{2l}}{2lx^{2l}},\qquad x\to\infty \label{eq:psi0a} \\
\psi_{1}(x)&=&\frac{1+2x}{2x^{2}}+\sum_{l=1}^{\infty}\frac{B_{2l}}{x^{2l+1}},\qquad x\to\infty, \label{eq:psi1a}
\end{eqnarray}
where the constant $B_k$ is the $k$-th Bernoulli number~\cite{Brychkov08}.

For the exact capacity under the Hilbert-Schmidt ensemble~(\ref{eq:CHS}), we now have in the limit~(\ref{eq:lim}),
\begin{eqnarray}\label{eq}
a_{0} &=& 1+o\left(\frac{1}{m}\right) \\
a_{1} &=& o\left(\frac{1}{m}\right) \\
a_{2} &=& -\frac{11}{4}+o\left(\frac{1}{m}\right)
\end{eqnarray}
and
\begin{eqnarray}\label{eq}
&&\psi_{1}(m+\alpha+1) = o\left(\frac{1}{m}\right) \\
&&\psi_{0}(m+\alpha+1)-\psi_{0}(\alpha+1) = -\psi_{0}(\alpha+1)+o\left(\ln m\right),
\end{eqnarray}
where only the needed orders of expansions have been displayed. The above results lead to
\begin{eqnarray}
\mathbb{E}_{\rm{HS}}\!\left[C\right] &=& \frac{\pi^2}{3}+\left(1+o\left(\frac{1}{m}\right)\right) o\left(\frac{1}{m}\right)+o\left(\frac{1}{m}\right) \\
&&\times\left(-\psi_{0}(\alpha+1)+o\left(\ln m\right)\right)-\frac{4}{11}+o\left(\frac{1}{m}\right),
\end{eqnarray}
where by using the fact that
\begin{equation}\label{eq}
\lim_{m\to\infty}\frac{\ln m}{m}=0
\end{equation}
one arrives at the asymptotic result~(\ref{eq:CHSa}),
\begin{equation}\label{eq}
\mathbb{E}_{\rm{HS}}\!\left[C\right] \stackrel{(\ref{eq:lim})}{\longrightarrow} \frac{\pi^2}{3}-\frac{4}{11}.
\end{equation}

Similarly, for the result under the Bures-Hall ensemble~(\ref{eq:CBH}), we obtain in the limit~(\ref{eq:lim}),
\begin{eqnarray}\label{eq}
b_{0} &=& 2+o\left(\frac{1}{m}\right) \\
b_{1} &=& o\left(\frac{1}{m}\right)
\end{eqnarray}
and
\begin{eqnarray}\label{eq}
&&\psi_{1}(m+\beta+1) = o\left(\frac{1}{m}\right) \\
&&\psi_{0}(m+\beta+1)-\psi_{0}(\beta+1) = -\psi_{0}(\beta+1)+o\left(\ln m\right).
\end{eqnarray}
Consequently, we have
\begin{eqnarray}
\mathbb{E}_{\rm{BH}}\!\left[C\right] &=& \frac{\pi^2}{3}+\frac{\pi^2}{3}+\left(2+o\left(\frac{1}{m}\right)\right)o\left(\frac{1}{m}\right)+o\left(\frac{1}{m}\right)\\
&&\times\left(-\psi_{0}(\alpha+1)+o\left(\ln m\right)\right)-\frac{\pi^2}{2}-1,
\end{eqnarray}
which leads to the claimed result~(\ref{eq:CBHa}),
\begin{equation}\label{eq}
\mathbb{E}_{\rm{BH}}\!\left[C\right] \stackrel{(\ref{eq:lim})}{\longrightarrow} \frac{\pi^2}{6}-1.
\end{equation}
This completes the proof of corollary~\ref{c}.

\section{Conclusion and outlook}\label{sec:con}
As an important step towards understanding the statistical distribution of entanglement capacity, we derived the exact and asymptotic average capacity formulas under the Hilbert-Schmidt and Bures-Hall ensembles in this work. Key ingredients in obtaining the results are the orthogonal polynomial systems of the underlying random matrices and the machinery to process the resulting summations of special functions. Future works include computing higher-order statistics such as the variance of entanglement capacity under different random density models. It will be also of interest to discover other variants of capacity, besides the annealed capacity, that are described by the same limiting behavior, satisfy the monotonicity property from separable to maximally-entangled states, or/and lead to closed-form formulas of capacity statistics.

\section*{Acknowledgments}
The author wishes to thank Youyi Huang and Kazumi Okuyama for correspondence. This work is supported in part by the U.S. National Science Foundation ($\#$2150486).

\appendix

\section{List of summation identities}\label{sec:app}
In this appendix, we list the summation identities that have been utilized in this work. Among other references, the identities can be found in~\cite{Wei17,Wei20,HWC21,Wei20c,HW22} except for the last one~(\ref{eq:Ab}), which was derived in~\cite[(B.17)]{Milgram}. Note that the identities of summations of the form~(\ref{eq:A1s}) are excluded here, which can be found, for example, in the appendices of~\cite{Wei20}. For the listed identities, it is sufficient for our purposes to assume that the parameters $a,b$ are non-negative real numbers and $\beta$ is a positive half integer.

\begin{equation}\label{eq:A1}
\fl\sum_{k=1}^{m}\frac{(n-k)!}{(m-k)!}=\frac{n!}{(m-1)!}\frac{1}{n-m+1}
\end{equation}
\begin{equation}\label{eq:A2}
\fl\sum_{k=1}^{m}\frac{(n-k)!}{(m-k)!}\frac{1}{k}=\frac{n!}{m!}\left(\psi_{0}\left(n+1\right)-\psi_{0}\left(n-m+1\right)\right)
\end{equation}
\begin{eqnarray}\label{eq:A3}
\fl\sum_{k=1}^{m}\frac{(n-k)!}{(m-k)!}\frac{1}{k^{2}}&=&\frac{n!}{m!}\sum_{k=1}^{m}\frac{\psi_{0}(k+n-m)}{k}+\frac{n!}{m!}\Bigg(\frac{1}{2}\Big(\psi_{1}(n-m+1)-\psi_{1}(n+1)+\nonumber\\
&&\psi_{0}^{2}(n-m+1)-\psi_{0}^{2}(n+1)\Big)+\psi_{0}(n-m)(\psi_{0}(n+1)-\psi_{0}(m+1)-\nonumber\\
&&\psi_{0}(n-m+1)+\psi_{0}(1))\Bigg)
\end{eqnarray}
\begin{equation}\label{eq:A4}
\fl\sum_{k=1}^{m}\frac{\psi_{0}(k+a)}{k+a}=\frac{1}{2}\left(\psi_{1}(a+m+1)-\psi_{1}(a+1)+\psi_{0}^{2}(a+m+1)-\psi_{0}^{2}(a+1)\right)
\end{equation}
\begin{eqnarray}\label{eq:Aab}
\fl\sum_{k=1}^{m}\frac{\psi_{0}(k+a)}{k+b}&=&-\sum_{k=1}^{m}\frac{\psi_{0}(k+b)}{k+a}+\psi_{0}(a+m+1)\psi_{0}(b+m+1)-\psi_{0}(a+1)\psi_{0}(b+1)+\nonumber\\
&&\frac{1}{a-b}\left(\psi_{0}(a+m+1)-\psi_{0}(b+m+1)-\psi_{0}(a+1)+\psi_{0}(b+1)\right)
\end{eqnarray}
\begin{eqnarray}\label{eq:Ab}
\fl\sum_{k=1}^{2\beta}\frac{\psi_{0}(k-\beta)}{k}&=&\frac{1}{2}\psi_{1}(\beta+1)+\psi_{0}(\beta+1)\left(\psi_{0}(2\beta+1)-\psi_{0}(1)\right)-\frac{3}{2}\psi_{1}(1)
\end{eqnarray}

\end{document}